# Peridynamic modeling of the crack velocity dependence via an incubation time fracture criterion

M. Ignatev*[a], P. Weißgraeber[b], E. Oterkus[c], L. Radtke[a]

[a] Chair of Structural Mechanics, Faculty of Mechanical Engineering and Marine Technologies, University of Rostock, Germany

[b] Chair of Lightweight Design, Faculty of Mechanical Engineering and Marine Technologies, University of Rostock, Germany

[c] PeriDynamics Research Centre, Department of Naval Architecture, Ocean and Marine Engineering, University of Strathclyde, UK

**Abstract**

This study investigates one of the central problems of dynamic fracture mechanics, namely the dependence of the instantaneous stress intensity factor (SIF) on the crack propagation velocity. For this purpose, the well-known experiments by Ravi-Chandar and Knauss on brittle, amorphous Homalite-100 polymer plates are modeled using a peridynamic approach. The numerical model integrates the previously proposed remote stress fracture criterion into an incubation time fracture criterion. Results of numerical modeling indicate a significant variation in SIF values at an almost constant crack propagation velocity. Moreover, for higher crack propagation velocities, micro-branching is obtained numerically, leading to a larger scatter of SIF values. These effects were also observed in the experiments of Ravi-Chandar and Knauss, which provides new insights into the nature of the crack-velocity dependence of the Mode-I SIF.

## 1. Introduction

Dynamic fracture continues to be an active research field, owing to its complex and still unresolved nature. The dynamic effects obtained in various studies contradict the classical ideas based on the mechanism of quasi-static fracture [1]. These include the presence of maximum crack propagation velocity of the material [2], crack arrest [3], branching [4], spallation [5], and fracture delay [6].

Over the years, various approaches have been developed to model dynamic fracture. On the analytical side, classical works within linear elastic fracture mechanics (LEFM) established the role of dynamic SIF and associated criteria for crack initiation and propagation [7]. To address the limitations of such singular-field formulations in practical computations, numerical

*m.ignatev@uni-rostock.de, Corresponding author

**Nomenclature**

| *Acronyms* | |
|---|---|
| SIF | stress intensity factor |
| LEFM | linear elastic fracture mechanics |
| ITFC | incubation time fracture criterion |
| FE | finite-element |
| TCD | theory of critical distances |
| FFM | finite fracture mechanics |
| RS | remote stress |
| DTCD | dynamic theory of critical distances |
| | |
| *Symbols* | |
| $K_I$ | Mode-I instantaneous stress intensity factor |
| $\dot{a}$ | crack propagation velocity |
| $t$ | time |
| $K_{Ic}$ | critical static stress intensity factor |
| $P$ | generalized load |
| $a$ | crack length |
| $K_I^d$ | dynamic Mode-I instantaneous stress intensity factor |
| $K_{ID}$ | critical dynamic stress intensity factor |
| $d$ | characteristic size (or length) in the incubation time fracture criterion |
| $\tau$ | incubation time |
| $\rho$ | mass density |
| **x** | position of a peridynamic point |
| **ü** | acceleration of a peridynamic point |
| **b** | body force density of a peridynamic point |
| **x**′ | position of a neighboring peridynamic point |
| $\mathcal{H}_\mathbf{x}$ | spherical neighborhood of the peridynamic point **x** |
| $\delta$ | radius of the spherical neighborhood $\mathcal{H}$, also called horizon |
| $V_{\mathbf{x}'}$ | volume occupied by the peridynamic element of point **x**′ |
| **T** | force vector state |
| **Y** | deformation vector state |

| $\sigma_{yy}$ | normal local stress on the plane with normal along the $y$-axis |
| --- | --- |
| $x'$ | spatial coordinate used for integration |
| $t'$ | time variable used for integration |
| $\sigma_c$ | ultimate tensile stress of the material |
| $r$ | distance from the crack tip |
| $\sigma_{ij}^{bond}$ | bond stress tensor |
| $\sigma_{ij}^{A}$ | local stress tensor at the peridynamic point A |
| $\alpha$ | angle between the normal to the considered plane and $y$-axis |
| $\sigma_{yy(\alpha)}$ | normal local stress on a plane with normal rotated by an angle $\alpha$ about $y$-axis |
| $\sigma_{yy(\alpha)}^{bond}$ | normal bond stress on a plane with normal rotated by an angle $\alpha$ about $y$-axis |
| $\Delta x$ | grid spacing of the peridynamic discretization |
| $E$ | Young's modulus |
| $\nu$ | Poisson's ratio |
| $f$ | external force |
| $L$ | characteristic length in the theory of critical distances |
| $\dot{Z}$ | loading rate in the dynamic theory of critical distances |
| $J$ | J-integral value |
| $W$ | strain energy density |
| $n_i$ | normal vector component |
| $w_i$ | auxiliary function component |
| $u_i$ | displacement vector component |
| $x_i$ | Cartesian coordinates ($x_1 = x$, $x_2 = y$) |
| $F_{ij}$ | deformation gradient component |
| $\delta_{ij}$ | Kronecker delta |

frameworks such as cohesive zone models [8], [9] and more recently phase-field models [10], [11] have been introduced, offering greater flexibility in handling complex crack topologies. Nevertheless, these approaches face challenges such as strong dependence on traction–separation laws or the need for an intrinsic length scale.

In the scope of this work, we use peridynamics [12, 13] to numerically investigate an important effect of dynamic fracture: the dependence of the SIF $K_I$ on the crack velocity $\dot{a}$. This $K_I - \dot{a}$ dependence was observed in a series of classical experiments by Ravi-Chandar and Knauss on brittle fracture in amorphous Homalite-100 polymer plates, representing an unbounded medium [14]. The dependence identified in those experiments was found to be non-unique, in the sense that for the same crack velocity different values of $K_I$ were observed, motivating the present investigation.

An early study applying peridynamics to model the SIF time dependence for dynamically initiated crack propagation in an unbounded body has been undertaken in [15], utilizing Griffith's fracture criterion. Other peridynamic studies do not address the present problem directly: they either analyse stationary cracks under dynamic loading without crack propagation, or they simulate dynamic crack propagation without determining the time history $K_I(t)$. In [15], a pressure increasing at a constant rate was applied to the faces of a pre-existing crack. As a result, the crack propagation velocity initially increased and subsequently reached a constant steady-state value. The computed SIF time dependence fluctuated slightly around a constant value, which was slightly higher than the critical static SIF $K_{Ic}$ (apparently due to inertia or stress wave effects which are naturally incorporated into the peridynamic modelling approach). Nevertheless, since Griffith's approach does not account for loading-rate effects, it is not capable of reproducing the experimentally observed dependence of the SIF $K_I$ on the crack velocity $\dot{a}$ mentioned above.

Within the broad spectrum of approaches to dynamic fracture, two approaches to dynamic fracture criteria can be distinguished. The first one, originating from the works of Freund [16-18] and later developed by Rosakis [19], is based on an assumption that fracture criterion can be written in the form:

$$K_I^d(P(t), a(t), \dot{a}(t)) = K_{ID}(\dot{a}(t)), \qquad (1)$$

with $K_I^d(P(t), a(t), \dot{a}(t))$ being the dynamic Mode-I SIF, depending on the generalized load, crack length, and crack velocity, and $K_{ID}(\dot{a}(t))$ denoting the critical dynamic SIF as a function of the crack velocity.

This approach can describe some of the experimentally observed phenomena of dynamic fracture, mainly in the case of well-developed plasticity (though still within the framework of small-scale yielding). The limitation, however, as mentioned by Owen [20], is that generally the left-hand side functional, $K_I^d(P(t), a(t), \dot{a}(t))$, can only be obtained numerically, which complicates the implementation of criterion (1) in practice. Moreover, determining the right-hand side function $K_{ID}(\dot{a}(t))$ over the entire range of crack velocities is an experimentally demanding task.

In this paper, we employ the incubation time fracture criterion (ITFC) [21], which introduces a characteristic fracture time to capture rate effects in brittle fracture. In [22], ITFC was implemented within a finite-element (FE) framework to reproduce the experimentally observed $K_I - \dot{a}$ dependence. While this study demonstrated that ITFC can qualitatively capture the observed trends, significant quantitative discrepancies remained, primarily due to the inherent limitations of local continuum mechanics in representing the singular crack-tip fields.

Conceptually, ITFC shares features with the theory of critical distances (TCD) [23], in the sense that both introduce a characteristic length $d$ to regularize the crack-tip singularity and to define a minimum fracture increment. However, unlike in TCD, where $d$ is an externally calibrated material length, the characteristic length in the classical formulation of ITFC by Petrov and Morozov arises directly from the quasi-static Irwin criterion (see Section 3.2) and ensures the consistency of the time-averaged stress condition with the classical fracture toughness. Thus, ITFC includes a length scale, but it is prescribed rather than intrinsic. In contrast, finite fracture mechanics (FFM) with the Coupled Criterion [24, 25] determines the finite crack advance intrinsically: the crack increment results from the coupled satisfaction of local stress and global energy requirements, without prescribing a fixed length in advance. This distinction also points to a possible future extension: combining the space–time framework of ITFC with the intrinsic, geometry- and loading-dependent length of FFM to achieve a fully dynamic and predictive fracture criterion. A comprehensive overview of FFM and the coupled stress–energy criterion, with a focus on crack initiation at singular and non-singular stress concentrations, is provided in the review by Weißgraeber et al. [26]. While this review primarily addresses quasi-static formulations, it also outlines open challenges and possible extensions of FFM toward dynamic loading conditions. Recent advances have extended Finite Fracture Mechanics (FFM) with the Coupled Criterion to dynamic loading scenarios by incorporating inertial effects into the energy balance and accounting for crack velocity during initiation [27-29]. These developments retain the intrinsic finite-length prediction of classical FFM while capturing the rate dependence of crack initiation without resorting to empirically rate-adjusted material properties.

In parallel to these developments, the present work embeds ITFC into a peridynamic formulation, thereby providing a nonlocal description of the evolving fracture process. Peridynamics, a nonlocal reformulation of continuum mechanics that is particularly well suited for modeling crack initiation and propagation, describes the motion of material points by summing the nonlocal forces between them, which does not entail any restrictions with regard to discontinuities such as cracks. To perform the present simulations, the peridynamic ITFC is formulated based on previously proposed remote stress (RS) fracture criterion [30]. The successful prediction of the critical SIF dependence on the crack initiation time using the peridynamic ITFC in [31, 33] (see Fig. 1) motivates its application to dynamic crack propagation modeling in the present study.

The remainder of the work is structured as follows: Section 2 briefly describes the experiments on the $K_I - \dot{a}$ for a dynamically propagating crack; Section 3.1 provides an overview of the peridynamic approach; Section 3.2 develops a peridynamic version of the ITFC; Section 3.3 details the dynamic crack propagation simulations; and Section 4 presents the results.

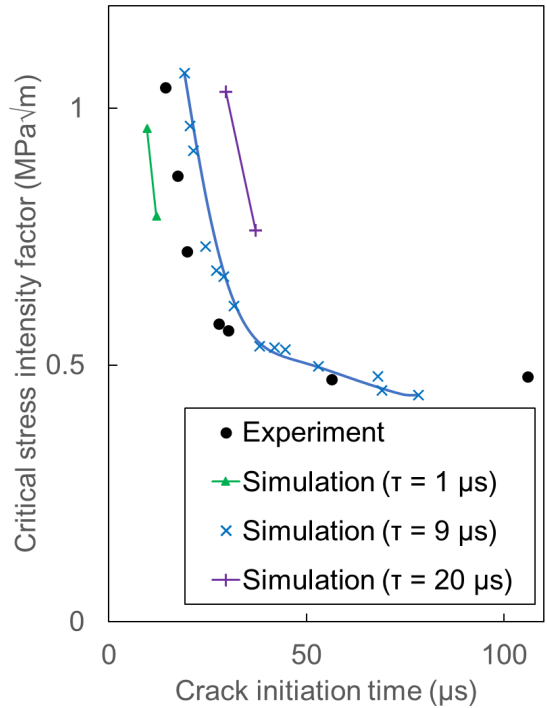

Fig. 1. Dependence of the critical Mode-I SIF on the crack initiation time. Black dots: experimental data of Ravi-Chandar and Knauss [34]. Green triangles: peridynamic simulations [31, 33] with ITFC for incubation time $\tau = 1\ \mu s$; blue crosses: $\tau = 9\ \mu s$; purple pluses: $\tau = 20\ \mu s$. Solid lines: averaged simulation results. For $\tau = 1\ \mu s$ and $\tau = 20\ \mu s$ only two representative points are shown because these $\tau$-values were tested only illustratively to demonstrate the horizontal shift of the dependence of the critical Mode-I SIF on the crack initiation time curve with varying incubation time.

## 2. Experimental $K_I - \dot{a}$ dependence

In the end of the last century, Ravi-Chandar and Knauss conducted a series of experiments on the dynamic crack initiation, propagation and crack arrest in the brittle samples made of a glassy polymer material (Homalite-100) [14, 34-36]. In [14], the $K_I - \dot{a}$ dependence was observed in the experiments on dynamic crack propagation. In these tests, a trapezoidal pulse with a 25 $\mu s$ rise time was applied to the pre-cracked surface of a sample using an electromagnetic loading device. By varying the maximum load achieved within 25 $\mu s$, the corresponding constant loading rate was obtained in each test. The sample had dimensions of approximately 300 mm x 500 mm (the size of the sample is omitted in [14] but given in [7]), that allows to consider it as an unbounded body during the first 150 $\mu$s from the onset of load application.

Using the caustic method, which determines the dynamic SIF based on the shape of caustics formed by the deflection of reflected light rays in the crack-tip region [37], and cinematography, the histories of the SIF and the crack extension were obtained. The absence of reflected elastic waves at the crack tip during the tests resulted in a constant crack propagation velocity throughout each test. Higher loading rates led to increased crack propagation velocities; however, significant fluctuations in instantaneous SIF values were still observed despite these constant velocities. Consequently, a specific crack propagation velocity does not correspond uniquely to a single SIF value; instead, a distinct range of SIF values is associated with each velocity.

In Fig. 2, the intervals of SIF values for each test from [14] (at the respective crack propagation velocity) are shown as the horizontal segments. Starting from a certain crack propagation velocity on, microbranching occurs, producing a more pronounced SIF range at higher velocities. Clearly, these results indicate the absence of one-to-one $K_I - \dot{a}$ relationship.

At the same time, the experimental results obtained by Ravi-Chandar and Knauss can be questioned as they contradict the results obtained by Dally [38] showing a unique $K_I - \dot{a}$ dependence (see Fig. 2). In their study, dynamic fracture experiments on Homalite-100 specimens using various geometrical configurations were performed. The method of photoelasticity was employed for determining the SIF at the crack tip there. The qualitative differences observed in the $K_I - \dot{a}$ relationships reported in the aforementioned experiments may stem from variations in the methods used to determine the instantaneous SIF. However, an investigation of this issue lies beyond the scope of the present study.

## 3. Theory and modeling

This section presents the theoretical background and numerical framework used in this study, including the peridynamic formulation, the incubation-time fracture criterion, and the modeling procedure for dynamic crack propagation.

### 3.1. Peridynamics

In the present work, a peridynamic approach [12, 13] is used to study the problem of dynamic crack propagation. In peridynamic theory, the body is represented as a set of discrete points governed by inherently non-local interaction rules. Within this framework, the medium balance equation is expressed without spatial differential operator, facilitating the representation of discontinuities such as cracks. An ordinary state based material model from [13] is used here, for which the equations of motion are as follows:

$$\rho(\mathbf{x})\ddot{\mathbf{u}}(\mathbf{x},t) = \int_{\mathcal{H}_\mathbf{x}} \left\{ \underline{\mathbf{T}}[\mathbf{x},t]\langle \mathbf{x}' - \mathbf{x}\rangle - \underline{\mathbf{T}}[\mathbf{x}',t]\langle \mathbf{x} - \mathbf{x}'\rangle \right\} dV_{\mathbf{x}'} + \mathbf{b}(\mathbf{x},t), \qquad (2)$$

where $\rho(\mathbf{x})$, $\ddot{\mathbf{u}}(\mathbf{x},t)$, $\mathbf{b}(\mathbf{x},t)$ are mass density, acceleration and body force density, respectively; $\mathbf{x}'$ is the position vector of any neighbor of the peridynamic point $\mathbf{x}$ in its spherical neighborhood $\mathcal{H}_\mathbf{x}$ of radius $\delta$ (named horizon); $V_{\mathbf{x}'}$ is the volume occupied by the peridynamic element of point $\mathbf{x}'$. $\underline{\mathbf{T}} = \underline{\mathbf{T}}(\underline{\mathbf{Y}})$ is the force vector state function, which maps the deformation vector state $\underline{\mathbf{Y}}$ into the force-vector state $\underline{\mathbf{T}}[\mathbf{x},t]$. In the angle brackets $\langle \mathbf{x}' - \mathbf{x}\rangle$ denotes the bond vector, i.e., the relative position vector between two interacting peridynamic points on which a force state operates.

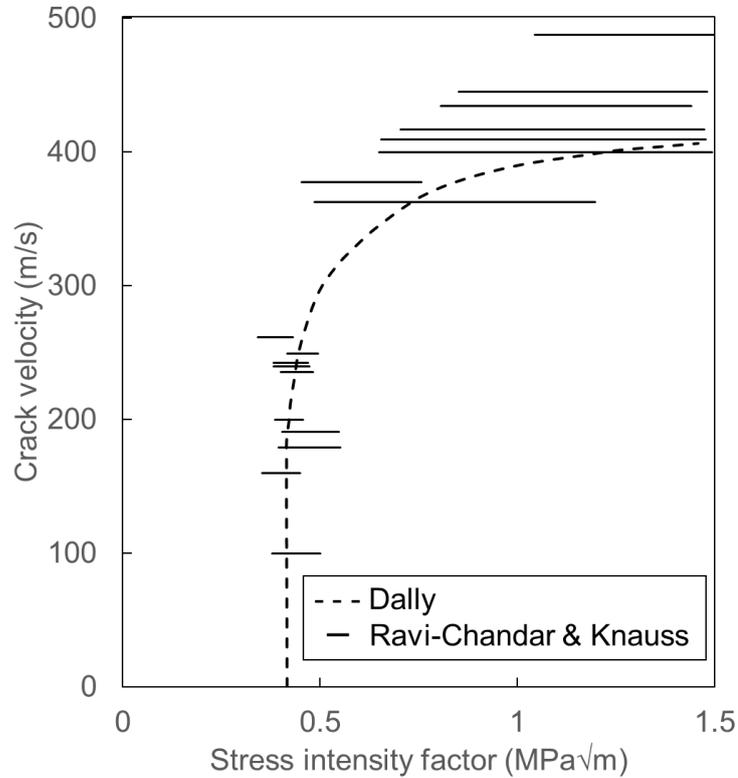

Fig. 2. Experimentally observed dependence of crack propagation velocity on the instantaneous Mode-I SIF in Homalite-100. The figure is redrawn from Fig. 5 in [14]. Black lines: data from Ravi-Chandar & Knauss [14]; dashed line: data from Dally [38]. Each horizontal segment represents the experimentally measured range of SIF values corresponding to a particular (near-constant) crack propagation velocity during dynamic fracture of Homalite-100.

Peridynamic simulations in this work are conducted in the open-source software Peridigm [39, 40] (implemented for 3D case), which is developed by Sandia National Laboratories. Due to the plane stress state, failure of the out-of-plane bonds in the 3D model is assumed to make a negligible contribution to the overall crack propagation.

### 3.2. Incubation time criterion of fracture

Since the classical approach of fracture mechanics considers a critical value for the SIF that is independent of the loading rate, it fails to predict the $K_I - \dot{a}$ dependence. At the same time, the dynamic version of the Irwin's criterion (1) inherently yields a unique $K_I - \dot{a}$ dependence. In view of the experimentally observed non-unique $K_I - \dot{a}$ dependence [14], the ITFC [21, 41, 42] is employed in this work. The ITFC accounts for the average value of the stress variations both in space, over a characteristic length $d$, and in time, over the characteristic fracture time $\tau$:

$$\frac{1}{\tau}\int_{t-\tau}^{t}\frac{1}{d}\int_{0}^{d}\sigma_{yy}(x',t')dx'dt' \geq \sigma_c, \tag{3}$$

where $t$ is the current time of the simulation; $\tau$ is an incubation time of fracture; $\sigma_{yy}(x',t')$ is the normal stress value at point with coordinate $x'$ along the axis of the crack line at time of integration $t'$; coordinate system is chosen such that the $x$-axis is horizontal and the $y$-axis is vertical (see Fig. 3); $\sigma_c$ is the ultimate tensile stress of the material. A characteristic size $d$ is determined from the condition of correspondence of the ITFC to the Irwin's criterion within the framework of LEFM:

$$d = \frac{2}{\pi}\frac{K_{Ic}^2}{\sigma_c^2}, \tag{4}$$

and this parameter is regarded as a material constant.

It is supposed that the fracture is not an instantaneous process, but on the contrary, a particular amount of time $\tau$ is needed for the preparatory processes (microcracking, coalescence of defects and pores, etc.) to develop and to cause the macroscopic material failure. The incubation time $\tau$ is introduced as a material constant and its value is determined from dynamic fracture experiments. ITFC has been applied in a number of studies to predict critical fracture parameters in both homogenous and heterogeneous materials [43, 44], and to model dynamic fracture processes such as impact, crack propagation, spallation, and solid particle erosion [45-48].

To embed the ITFC into a peridynamic framework, we build upon the previously proposed RS fracture criterion [30], which provides a convenient basis for expressing the bond-level failure condition. To this end, let us first recall the concept of the bond stress tensor $\sigma_{ij}^{bond}$:

$$\sigma_{ij}^{bond} = \frac{\sigma_{ij}^A + \sigma_{ij}^B}{2}, \tag{5}$$

where $i,j = 1..3$ and the local stress tensors $\sigma_{ij}^A$ and $\sigma_{ij}^B$ are defined at the peridynamic points A and B respectively. In a quasi-static case, where inertia effects are negligible, we set $\tau = 0$ and the outer integral in relation (3) vanishes:

$$\frac{1}{d}\int_0^d \sigma_{yy}(x')dx' \geq \sigma_c. \tag{6}$$

In view of the Sneddon-Williams asymptotic for a Mode-I crack along the crack line at a distance $r$ from the crack tip:

$$K_I \approx \sigma_{yy}(r)\sqrt{2\pi r}, \tag{7}$$

condition (6) yields the classical Irwin's criterion [31] (see **Appendix A**):

$$K_I \geq K_{Ic}. \tag{8}$$

Substituting (7) into (8), we obtain:

$$\sigma_{yy}(r) \geq \frac{K_{Ic}}{\sqrt{2\pi r}}. \tag{9}$$

For analyzing crack propagation in an arbitrary direction in 2D, one considers the normal stress components $\sigma_{yy(\alpha)}(r)$ on planes with normal rotated by an angle $\alpha$ about $y$-axis. On the other hand, the bond stress $\sigma_{yy(\alpha)}^{bond}$ corresponds to the local stress $\sigma_{yy(\alpha)}(r)$, evaluated at a finite distance $r$ from the crack tip [30]. Thus, for a uniform discretization with grid spacing $\Delta x$ along the coordinate axes, relation (9) yields the following critical bond stress condition:

$$\sigma_{yy(\alpha)}^{bond} \geq \frac{K_{Ic}}{\sqrt{2\pi\Delta x}}. \tag{10}$$

Relation (10) is a 2D version of RS criterion from [30]. Here, we provide a simplified derivation of the RS criterion that does not involve the characteristic length $d$ (in contrast to [30]). Introducing $d$ into the RS criterion is unnecessary when modeling crack propagation from a pre-crack, since the ultimate strength $\sigma_c$ does not affect the propagation condition.

In the dynamic case, the ITFC for bond failure can be formulated by integrating (10) over the incubation time $\tau$. This formulates the peridynamic version of Eq. (3):

$$\frac{1}{\tau}\int_{t-\tau}^{t} \sigma_{yy(\alpha)}^{bond}(t)dt \geq \frac{K_{Ic}}{\sqrt{2\pi\Delta x}}. \tag{11}$$

Note that Eq. (11) is equivalent to the ITFC for bond failure used in [31, 33]. In this work, $\sigma_{yy(\alpha)}^{bond}$ in Eq. (11) is evaluated for $\alpha = 0°$ and $\alpha = 90°$, i.e. for planes parallel and perpendicular to the crack axis (in [31, 33] only for $\alpha = 0°$). This restriction allows the crack to propagate along the horizontal $x$-axis with possible microbranching, while suppressing full branching of the main crack into arbitrary directions $0° < \alpha < 90°$.

In the numerical implementation, the integration in Eq. (11) is performed using the trapezoidal rule at each time step. For every peridynamic point, the stress history from the preceding time steps is stored and updated, and the number of stored steps corresponds to the temporal window equal to the incubation time $\tau$.

In the case of arbitrary 3D crack propagation, the stress values $\sigma_{yy(\alpha,\beta,\gamma)}^{bond}(t)$ in Eq. (11) (instead of $\sigma_{yy(\alpha)}^{bond}(t)$) should be evaluated on multiple planes rotated by angles $\alpha, \beta, \gamma$ about the $x, y$ and $z$-axes, respectively, as done in [30] for the RS criterion (10).

### 3.3. Dynamic crack propagation modeling

In the present study, the crack propagation experiments of Ravi-Chandar and Knauss [14] are modeled to compute the $K_I - \dot{a}$ dependence for dynamically initiated crack propagation. In the simulations, the specimen dimensions is chosen identical to those used in the experiments, i.e., 300 mm × 500 mm with a 50 mm pre-notch. With this size, elastic waves reflected from the specimen boundaries do not reach the crack tip within the first 150 $\mu$s.

The characteristic lengths in the discretization are chosen as $\Delta x$ = 0.5 mm and $\delta = 3\Delta x$. The specimen thickness is equal to the diameter of the spherical horizon $2\delta = 3mm$. The mechanical properties of Homalite-100 used in the calculations are summarized in Table 1. The characteristic size $d$ is obtained from Eq. (4) using the values of $K_{IC}$ and $\sigma_c$.

In [21] the incubation time $\tau$ is determined by fitting an analytical expression that relates the critical SIF to the crack initiation time for the relevant experimental loading rates. By adjusting the incubation time $\tau$, the dependence of the critical SIF on the crack initiation time shifts horizontally, as illustrated in Fig. 1, while maintaining its characteristic shape. Thus, by fitting the analytical curve to the experimental data from [34], the incubation time $\tau$ was determined in [21].

The dynamic loading is realized via a volumetric load that is applied in a zone of a width $2\delta$ adjacent to the crack surface, excluding the crack tip vicinity reserved for the J-integral calculation (see Fig. 3). The force pulse has a trapezoidal shape with a rise time of 25 $\mu$s. The vertical normal stress rate is evaluated at a peridynamic point located within the region where the force is applied. The loading rates in the simulations vary from $3 \cdot 10^4$ MPa/s to $1.3 \cdot 10^5$ MPa/s, which is within the range of the experimental loading rates from $2.5 \cdot 10^4$ MPa/s to $6.2 \cdot 10^5$ MPa/s [14].

Table 1. Mechanical properties of Homalite-100 [49].

| Density, $\rho$ | 1230 kg/m$^3$ |
|---|---|
| Young's modulus, $E$ | 3.9 GPa |
| Poisson's ratio, $\nu$ | 0.35 |
| Critical SIF, $K_{IC}$ | 0.48 MPa$\sqrt{m}$ |
| Ultimate tensile stress, $\sigma_c$ | 48 MPa |
| Characteristic size, $d$ | 0.064 mm (Eq. (4)) |
| Incubation time, $\tau$ | 9 $\mu s$ [21] |

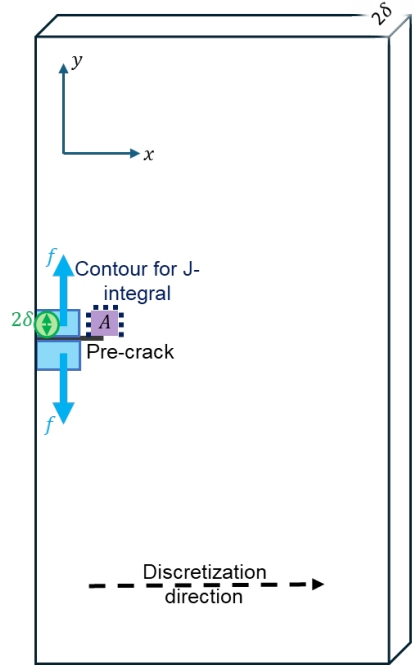

Fig. 3. Geometry, loading and boundary conditions for the peridynamic simulation of dynamic crack propagation. A 300 mm × 500 mm rectangular specimen with a 50 mm pre-crack is subjected to external forces $f$, applied symmetrically on the upper and lower boundaries to generate a constant loading rate. The figure indicates the discretization direction and the rectangular contour used for evaluating the dynamic J-integral. The contour encloses a domain $A$ surrounding the crack tip, ensuring that only the local peridynamic neighborhood $\mathcal{H}_x$ of the crack tip contributes to the SIF evaluation.

The dynamic stress intensity factor $K_I$ is computed using the dynamic J-integral following the formulation of Eq. (3.1b) in [50]. In the present work, a rectangular integration contour moving with the crack tip is used. A detailed derivation of the discrete form of the dynamic J-integral and the expressions employed in the peridynamic implementation are provided in **Appendix B**.

All simulations were carried out on a workstation equipped with an Intel(R) Core(TM) i7-4960X CPU @ 3.60 GHz and 32 GB RAM. The computations were performed using 6 parallel threads. The discretization consisted of 3 600 000 peridynamic points. An explicit time integration scheme was employed in all simulations. A time step of $\Delta t = 0.18\ \mu s$ was selected such that it satisfies the stability requirement of the explicit integration scheme and is an integer subdivision of the incubation time $\tau = 9\ \mu s$. With the chosen time step, the stress history used in the evaluation of Eq. (11) corresponds to the most recent 50 time steps (i.e., the interval $\tau$). Depending on the duration of crack propagation in each case, the wall-clock time of a single simulation ranged from approximately 84 to 135 hours.

## 4. Results and discussion

Fig. 4 and Fig. 5 present the time histories of the SIF and crack extension obtained in peridynamic simulations, compared with the corresponding experimental data. The simulation results show that, after an initial acceleration phase, a relatively constant crack propagation velocity is obtained in each case (as also reported in [15]). This contrasts with the experimentally observed crack propagation velocity, which remains constant from the very beginning of crack propagation. Fig. 4 shows the results for crack propagation velocities of 366 m/s and 328 m/s, with a loading rate of $6.6 \cdot 10^4$ MPa/s. Similarly, Fig. 5 shows the results for crack propagation velocities of 430 m/s and 445 m/s, obtained at a loading rate of $1.27 \cdot 10^5$ MPa/s. In each case, the minimum ("Min") and maximum ("Max") values of the SIF from the beginning of crack propagation ("Start") are identified. Comparisons between the numerical and experimental dependencies in Fig. 4 and Fig. 5 are presented for those cases where the modeled crack propagation velocity most closely matches the experimentally measured values.

The modeled SIF histories qualitatively mirror the experimentally measured dependencies, although there is some divergence. However, the authors of [14] caution that the absolute magnitudes of the experimentally obtained SIF histories may require adjustment. Note that the experimental SIF histories in Fig. 4-Fig. 5 do not start from zero, which contradicts the fact that the loading of the specimen starts at zero time.

The SIF histories presented in Fig. 4-Fig. 5 are not filtered in order to show the true nature of the SIF history. Of course, filtering (averaging) the SIF dependencies would significantly reduce their scatter, but they would still noticeably vary after the crack initiation similar to the experiments. Only once the crack velocity stabilizes, also the averaged SIF stays almost constant. This is in line with the experiments for the configuration from Fig. 4. However, for the configuration considered in Fig. 5, the SIF in the experiment stabilizes after a much longer time.

Fig. 6 compares the crack-velocity dependence of the SIF obtained from experiments and from the present peridynamic simulations. Note that each modeled SIF interval reflects the variations immediately following crack initiation, i.e., the initial acceleration phases are included. This corresponds to the limits indicated by "Min" and "Max" in Fig. 4-Fig. 5. As shown in Fig. 6, for intermediate crack velocities of about 100–300 m/s, the experimental SIF intervals range from approximately 0.34 to 0.55 MPa√m, while the numerical results give a similar spread of 0.48–0.79 MPa√m. At higher velocities (300–500 m/s), the experimental range broadens up to 0.45–1.5 MPa√m, whereas the simulations predict 0.53–1.25 MPa√m. These comparisons indicate that the model captures both the increase and the scatter of the $K_I - \dot{a}$ dependence reasonably well.

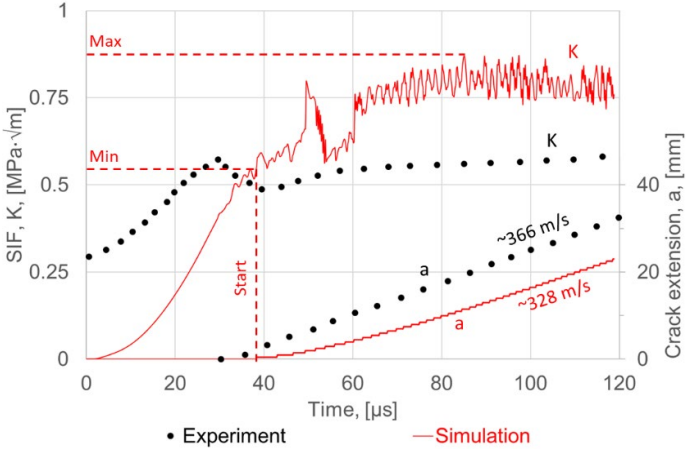

Fig. 4. Time histories of the Mode-I SIF $K$ and crack extension $a$: experiment vs. present peridynamic simulations for lower crack velocities. Black dots show the experimental data for a crack propagation velocity of 366 m/s in Homalite-100 [14]. Red curves show the corresponding peridynamic simulation results obtained for a crack propagation velocity of 328 m/s under a loading rate of $6.6 \cdot 10^4$ MPa/s. "Start" marks the onset of crack propagation as predicted by the ITFC, while "Min" and "Max" indicate the minimum and maximum SIF values recorded after crack initiation.

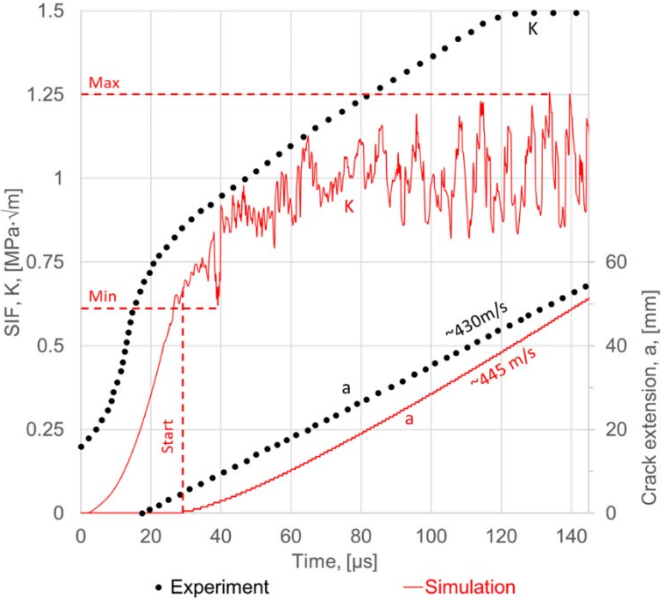

.
Fig. 5. Time histories of the Mode-I SIF $K$ and crack extension $a$: experiment versus the present peridynamic simulation at higher crack velocities. Black dots show the experimental data for a crack propagation velocity of 430 m/s in Homalite-100 [14]. Red curves show the corresponding peridynamic simulation results obtained for a crack propagation velocity of 445 m/s under a loading rate of $1.27 \cdot 10^5$ MPa/s. "Start" marks the onset of crack propagation as predicted by the ITFC, while "Min" and "Max" indicate the minimum and maximum SIF values recorded after crack initiation.

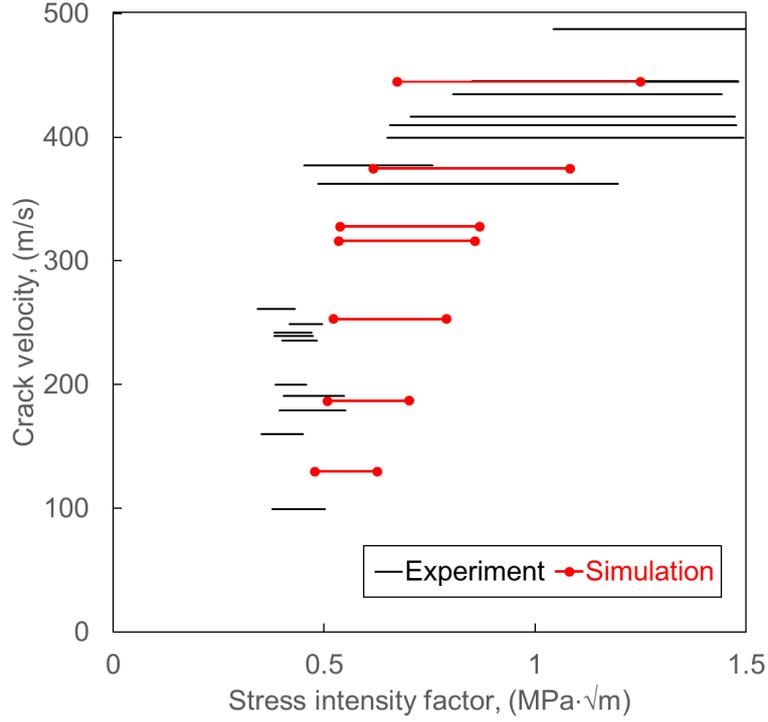

Fig. 6. Comparison of the experimentally observed and numerically predicted $K_I - \dot{a}$ dependencies. Black horizontal segments: experimental data from Ravi-Chandar & Knauss [14], obtained for loading rates ranging from $2.5 \cdot 10^4$ MPa/s to $6.2 \cdot 10^5$ MPa/s. Red segments: present peridynamic simulations for loading rates from $3 \cdot 10^4$ MPa/s to $1.3 \cdot 10^5$ MPa/s. Each segment represents the range of SIF values recorded during the crack propagation at the corresponding velocity.

Our investigations show that for crack propagation velocities above 400 m/s, the formation of microdefects near the main crack becomes noticeable (see Fig. 7). This is most likely the reason why the amplitude of the SIF oscillations for a velocity of 430 m/s in Fig. 5 is much larger than that at 328 m/s in Fig. 4, where no microbranching occurs. The onset of microbranching in the simulations (at velocities above 400 m/s) approximately corresponds to the experimentally observed onset of crack branching at about 400 m/s (see Fig. 8 in [14]).

Importantly, in every case, the moving contour used for the J-integral calculation exits the microbranching zone before it grows significantly. Consequently, the microbranches remain enclosed within the contour throughout the entire duration of the J-integral evaluation within that region (see Fig. 8).

As mentioned in Section 3.2, in the present study the ITFC (11) is evaluated only on two planes oriented at 0° and 90° with respect to the main crack axis. This restriction was introduced intentionally to suppress fully developed crack branching and to capture only local microbranching, while keeping the evaluation of the dynamic J-integral computationally feasible: a crack branching in arbitrary directions would otherwise require repeated recalculation of the contour around each new tip. It should be noted that this restriction is not expected to significantly affect the evaluation of the dynamic J-integral, as the contour is always constructed around the main crack tip. Possible contributions from the character and direction of development of microbranches are thus excluded as long as they remain outside the integration contour.

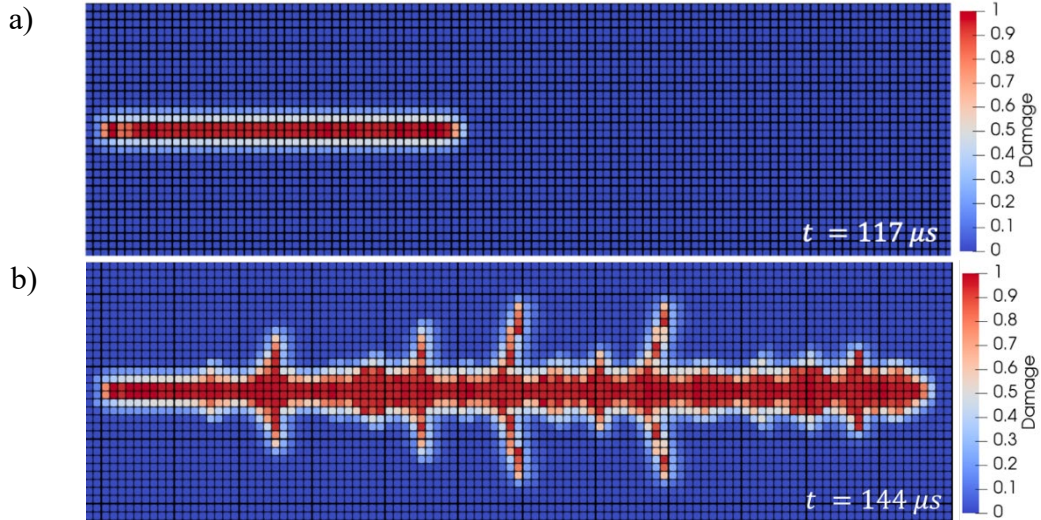

Fig. 7. Crack propagation patterns obtained from simulations. (a) Crack propagating without microbranching at a velocity of 375 m/s (final state at $t = 117\ \mu s$), obtained under a loading rate of $6.6 \cdot 10^4$ MPa/s. (b) Crack propagating with pronounced microbranching at a velocity of 445 m/s (final state at $t = 144\ \mu s$), obtained under a loading rate of $1.27 \cdot 10^5$ MPa/s. Damage values (0–1) represent the degree of failure of peridynamic points (with 0 = intact and 1 = fully broken).

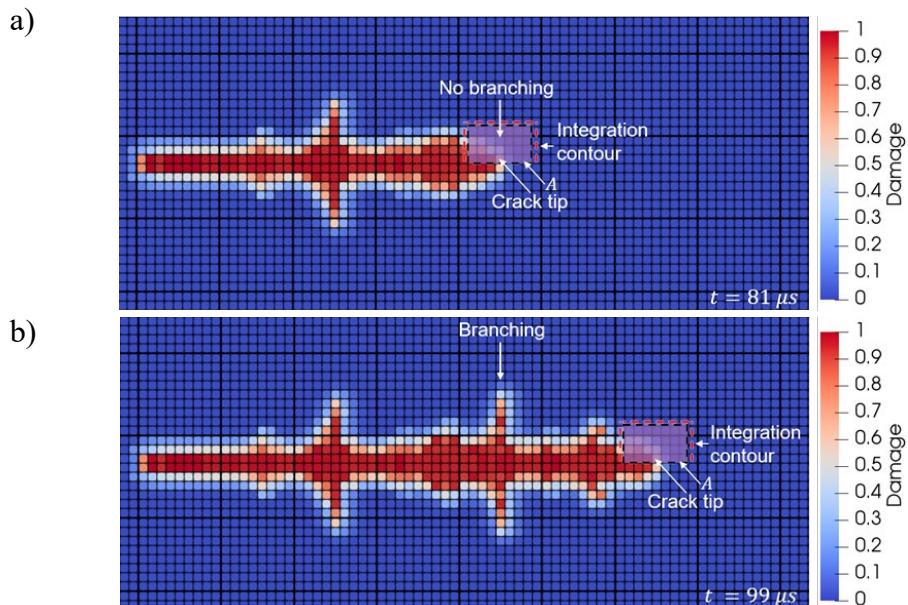

Fig. 8. Snapshots of crack propagation with the moving integration contour used for the dynamic J-integral calculation. (a) $t = 81\ \mu s$: no branching occurs; the contour encloses the main (unbranched) crack tip, which remains fully inside domain $A$. (b) $t = 99 \mu s$: microbranching occurs behind the main crack tip, while the contour (advancing with the main crack) excludes the branching region, ensuring a stable evaluation of the J-integral. Both snapshots correspond to the simulation with crack velocity 445 m/s under a loading rate of $1.27 \cdot 10^5$ MPa/s.

The appearance of microbranching in the numerical results suggests that the crack tip energy exceeds the threshold for crack branching observed experimentally. Therefore, the numerically obtained oscillations of the SIF appear physically justified, supporting that the $K_I - \dot{a}$ dependence is not unique.

The peridynamic RS criterion has been validated for cracks propagating in arbitrary directions [30]. Accordingly, evaluating $\sigma_{yy(\alpha)}^{bond}(t)$ on multiple planes with $0° < \alpha < 90°$ in criterion (11) enables the modeling of crack branching, and its combination with ITFC may produce a numerical SIF history in closer agreement with experiments.

To sum up, presented numerical simulations predict SIF histories similar to the experimental ones for dynamic crack propagation with constant velocity. However, experimental crack velocities reach their constant values instantaneously, while in the numerical simulations this stabilization requires a significant transient period of approximately 30–40 μs. At higher crack propagation velocities, the numerical simulations exhibit crack microbranching, which markedly amplifies the scatter in the SIF history. A thorough investigation of crack branching using the proposed model for arbitrary propagation directions should be undertaken in future work.

As discussed in this Section, our numerical results reveal several mechanisms that govern whether the experimentally observed $K_I - \dot{a}$ dependence is unique or not. The main findings are summarized below:

- The duration of the crack acceleration phase before the crack reaches its steady propagation velocity, and the corresponding evolution of the SIF during this interval.
- The magnitude and character of the SIF fluctuations about its approximately constant average value, and the influence of these fluctuations on the resulting $K_I - \dot{a}$ response.
- The abrupt rise in SIF at the instant of crack branching, and the impact of this jump on the overall $K_I - \dot{a}$ behavior.

The last aspect concerns whether the SIF values at the instant of branching characterize the original crack or the individual branches that form thereafter.

Note, that ITFC introduces a characteristic incubation time $\tau$ and evaluates failure by temporal averaging of the stress (or energy release rate) over the interval $[t - \tau, t]$. This formulation captures the rate-sensitivity of brittle fracture but requires careful calibration of $\tau$. A modified version was later proposed in [51], which introduced a two-parameter criterion combining a threshold for the maximum principal stress with damage accumulation governed by the hydrostatic stress. In contrast to the classical formulation, the integration starts when the threshold is exceeded and continues until rupture, which provides a closer link to damage kinetics and enables quantitative reproduction of complex fracture patterns under impact. Such differences highlight that while ITFC-type criteria can qualitatively capture the experimentally observed rate effects, the choice of variant directly influences predictive capability and the level of experimental calibration required.

At the same time, it should be noted that although the classical ITFC does not explicitly incorporate damage evolution, in the peridynamic formulation this effect is implicitly present: bonds are progressively broken once the time-averaged criterion is satisfied, which naturally introduces a damage zone around the crack tip.

It should be recalled that criterion (11) used in the present work is based on the RS fracture criterion. The RS criterion relies on a characteristic fracture length $d$, which is obtained by enforcing equivalence between the ITFC and the quasi-static fracture toughness $K_{Ic}$. In criterion (11), inertial effects are accounted for through the introduction of the incubation time $\tau$, whereas the characteristic length $d$ remains constant. By contrast, in the work of [52], the authors introduced a dynamic reformulation of the TCD (DTCD), where the characteristic length $L$ of the TCD (an analogue of the characteristic length $d$ in ITFC formulation) was defined as explicitly dependent on the loading rate $\dot{Z}$. However, in the study [53], dealing with dynamic fracture of unreinforced concrete, the characteristic length $L(\dot{Z})$ was found to be only marginally affected by the loading rate. Accordingly, the use of a constant $d$ in the present work does not represent an inconsistency, although a rate-dependent generalization of $d$ could be considered in future investigations.

## 5. Conclusion

In this study we employ an incubation time fracture criterion together with a peridynamic theory to numerically investigate the $K_I - \dot{a}$ relationship during dynamically initiated crack propagation in a brittle medium. The results presented in this paper yield an approximately constant crack propagation velocity concurrent with pronounced fluctuations in the SIF. Accordingly, the $K_I - \dot{a}$ dependence is not unique when modeled using the proposed discretization approach. The peridynamic implementation of incubation time approach is found to be a valuable approach for the assessment of the dynamic fracture toughness of a pre-cracked specimen and for describing the dynamic crack propagation over a significant range of loading rates.

## 6. Acknowledgements

The authors thank Prof. Yu. Petrov and Dr. N. Kazarinov for very helpful discussions. The authors are grateful to Mikhnovich I.V. for assistance in writing the program code.

## 7. Appendix A. Quasi-static reduction of the ITFC and recovery of the Irwin criterion

Starting from the quasi-static version of the ITFC (6):

$$\frac{1}{d}\int_0^d \sigma_{yy}(x')dx' \geq \sigma_c, \qquad \text{A. 1}$$

and using the Sneddon–Williams asymptotic for a Mode-I crack (7),

$$\sigma_{yy}(r) \approx \frac{K_I}{\sqrt{2\pi r}}, \qquad \text{A. 2}$$

the integral in (A. 1) evaluates to:

$$\frac{1}{d}\int_0^d \sigma_{yy}(x')dx' = \frac{1}{d}\frac{K_I}{\sqrt{2\pi}}\int_0^d \frac{dx'}{\sqrt{x'}} = \frac{1}{d}\frac{K_I}{\sqrt{2\pi}}[2\sqrt{x'}]_0^d = K_I\sqrt{\frac{2}{\pi d}}. \qquad \text{A. 3}$$

Hence, the quasi-static ITFC condition (A. 1) becomes:

$$K_I \geq \sigma_c \sqrt{\frac{\pi d}{2}}. \qquad \text{A. 4}$$

Using the expression for $d$ given in Eq. (4), which ensures consistency with the Irwin criterion,

$$d = \frac{2}{\pi}\frac{K_{Ic}^2}{\sigma_c^2}, \qquad \text{A. 5}$$

condition A. 4 immediately yields:

$$K_I \geq K_{Ic}. \qquad \text{A. 6}$$

## 8. Appendix B. Discrete formulation of the dynamic J-integral used in the simulations

In the present simulations, the SIF is obtained from the dynamic J-integral (Eq. 3.1b in [50]), which, in analogy to the displacement formulation of the classical J-integral [54], can be rewritten as:

$$J \approx 2\Delta x \sum_{\alpha=I}^{III}\sum_{k=1}^{m}\left[\frac{E}{1-\nu^2}\left(\frac{1}{2}Wn_1 - w_i\right) + \frac{1}{2}\rho\left(\frac{\partial u_i}{\partial t}\right)^2 n_1\right]$$
$$+ 2(\Delta x)^2 \sum_{s=1}^{q}\left(\rho\frac{\partial^2 u_i}{\partial t^2}\frac{\partial u_i}{\partial x_1} - \rho\frac{\partial u_i}{\partial t}\frac{\partial^2 u_i}{\partial x_1 \partial t}\right), \qquad \text{B. 1}$$

$\alpha$ denotes the segment index along the integration contour (see Fig. 9); $k$ is the index of a peridynamic point along the contour; $m$ is the number of points on the corresponding segment $\alpha$; $W$ is the strain energy density, defined as:

$$W = \left(\frac{\partial u_1}{\partial x_1}\right)^2 + \left(\frac{\partial u_2}{\partial x_2}\right)^2 + 2\nu\frac{\partial u_1}{\partial x_1}\frac{\partial u_2}{\partial x_2} + \frac{1-\nu}{2}\left(\frac{\partial u_1}{\partial x_2} + \frac{\partial u_2}{\partial x_1}\right)^2; \qquad \text{B. 2}$$

$w_I, w_{II}, w_{III}$ are auxiliary functions corresponding to the right, top, and left segments of the integration contour, respectively (see Eqs. (18a-c) in [54]), and defined as follows:

$$w_I = \left(\frac{\partial u_1}{\partial x_1}\right)^2 + v\frac{\partial u_1}{\partial x_1}\frac{\partial u_2}{\partial x_2} + \frac{1-v}{2}\left[\frac{\partial u_1}{\partial x_2}\frac{\partial u_2}{\partial x_1} + \left(\frac{\partial u_2}{\partial x_1}\right)^2\right],$$

$$w_{II} = \frac{1-v}{2}\frac{\partial u_1}{\partial x_1}\frac{\partial u_1}{\partial x_2} + \frac{1+v}{2}\frac{\partial u_1}{\partial x_1}\frac{\partial u_2}{\partial x_1} + \frac{\partial u_2}{\partial x_1}\frac{\partial u_2}{\partial x_2},$$

$$w_{III} = -w_I;$$

B. 3

$n_1$ is the first component of the outward unit normal vector of the contour; $u_i$ is the component of the displacement vector of the $k$-th peridynamic point; for convenience here, the coordinates are alternatively written in index notation as $x_1 = x$ and $x_2 = y$, where $x$ is a horizontal axis; $t$ is the time; $F_{ij}$ are the components of the deformation gradient of the $k$-th peridynamic point; $\delta_{ij}$ is the Kronecker delta; the deformation gradient $\frac{\partial u_i}{\partial x_j}$ is expressed as:

$$\frac{\partial u_i}{\partial x_j} = F_{ij} - \delta_{ij};$$

B. 4

$s$ denotes the index of a peridynamic point from domain $A$ inside the integration contour comprising $q$ peridynamic points.

A rectangular contour is drawn around the current location of the crack tip so that only neighborhood $\mathcal{H}_\mathbf{x}$ of the peridynamic point $\mathbf{x}$ located at the crack tip is inside the contour of the dynamic J-integral (Fig. 9). The contour moves together with the crack tip throughout the simulation to ensure a consistent evaluation of the dynamic J-integral. The SIF is then calculated using the relation for plane stress under Mode-I loading:

$$K_I = \sqrt{EJ}.$$

B. 5

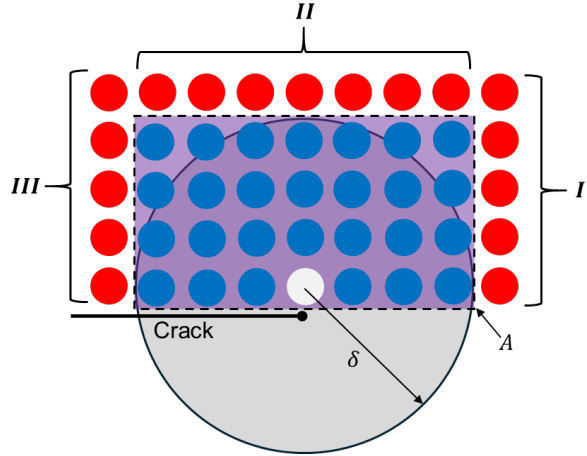

Fig. 9. Discretized rectangular contour used for the calculation of the dynamic J-integral. Red circles indicate peridynamic points located on the integration contour, while blue circles denote the points from domain $A$ inside the contour. The contour is divided into three segments ($I$, $II$, $III$) around the crack tip, along which the integration in B. 1 is performed. The contour is moved along with the crack tip throughout the simulation.